\documentclass[twocolumn]{aastex631}

\received{XXX}
\revised{YYY}
\accepted{ZZZ}

\shorttitle{Lowest X-ray state Spectrum}
\shortauthors{Kushwaha et al.}

\begin{document}

\title{On the Spectral Changes of OJ 287: The Lowest X-ray state spectrum-- extended at Optical-UV and Hard at X-rays}

\correspondingauthor{Pankaj Kushwaha}
\email{pankaj.kushwaha@iisermohali.ac.in} 

\author[0000-0001-6890-2236]{Pankaj Kushwaha}
% \author{Pankaj Kushwaha}\thanks{Email: pankaj.tifr@gmail.com}
\altaffiliation{DST-INSPIRE Faculty Fellow}
\affiliation{Department of Physical Sciences, Indian Institute of Science Education and Research Mohali, Knowledge City, Sector 81, SAS Nagar, Punjab 140306, India} 

\begin{abstract}
\noindent
Optical-UV synchrotron spectrum has been argued to be the primary driver of the
majority of X-ray spectral changes in the BL Lacetrae object OJ 287 during its
low and intermediate X-ray flux state. Here, we focus on the lowest recorded X-ray
flux state of OJ 287 by the Swift facility and report the finding of a power-law
optical-UV spectrum with a photon spectrum of $\rm 2.71\pm0.03$ continuing into
X-rays. Considering this at X-rays, we found a power-law X-ray spectrum of photon
spectral index $\rm 1.22\pm0.20$ that improves to $\rm 1.29\pm0.06$ when considering
other observations with similar X-ray spectra. This is the hardest reported X-ray
spectrum (0.3-10 keV) and is consistent with the reported Swift-BAT hard X-ray spectrum.
We further show that this X-ray spectrum can reproduce most of the flat X-ray spectra
when combined with the corresponding optical-UV continuum during the low and intermediate flux states strengthening synchrotron as the primary driver of most of the X-ray spectral
changes in the LBL state of the source. Combined with sharp-steepening/cutoff of the
optical-UV spectrum during bright phases, the extended-spectrum indicates a comparatively larger emission region and could be the large-scale jet emission. The optical-UV spectrum implies a high-energy power-law particle spectrum of $\rm \sim4.4$ while X-ray implies a hard low-energy particle spectrum of $\rm 1.3-1.6$ which could be the real or can result from a higher low-energy cut-off in the particle spectrum.
\end{abstract}

\keywords{galaxies: active -- BL Lacertae objects: general -- quasars: individual -- BL Lacertae objects: individual: OJ 287}

\section{Introduction}
\noindent
OJ 287 is an optically bright BL Lacertae (BLL) type object at a cosmological redshift
of z=0.306, characterized by a non-thermal continuum-dominated optical spectrum
with very weak emission line features reported only during its very faint optical
brightness states \citep{1985PASP...97.1158S,2010A&A...516A..60N,2021ApJ...920...12H}.
Together with flat spectrum radio quasars (FSRQs), they are referred to as blazars --
active galactic nuclei hosting large-scale powerful relativistic jets directed
roughly towards the Earth. Blazars are characterized by a highly variable continuum
extending across the entire electromagnetic (EM) spectrum from radio to GeV-TeV
gamma-rays that exhibits a broad, bi-model spectral energy distribution \citep[e.g.][]
{2010ApJ...716...30A,2015ApJ...807...79H,2022JApA...43...79K}. The temporal continuum variation is  primarily erratic and variable at all times from decades and more
to all the way to minutes scales accessible within the sensitivities of currently 
existing facilities \citep[e.g.][]{2018ApJ...863..175G}.

The low-energy emission component of the bi-model broadband SED that starts at
radio and peaks in between near-infrared (NIR) to X-ray energies is widely accepted
to be the synchrotron emission from relativistic electrons within the jet due to the
non-thermal nature of radio and optical spectrum together with their strong and variable polarization. The high-energy part is debated to be either via inverse Comptonization (IC)
-- a natural and logical scenario given relativistic electrons and strong photon
fields, or hadronic processes or a combination of both \citep[e.g.][and references therein]{2018NatAs.tmp..154G,2019ARNPS..69..477M}. Claims of detection of neutrinos from the direction of these sources support hadronic component and modeling indicates
the hadronic component to be sub-dominant at MeV-GeV energies \citep{2018NatAs.tmp..154G,2019ARNPS..69..477M}. 

A remarkable property of the bi-modal broadband SED is that the frequency at which
the low-energy emission component peaks rarely changes despite the observed strong
flux variations, most of which are often accompanied by a change in the continuum
spectrum in different observational bands. This stability combined with the physical understanding of the low-energy part of the SED has led to a physical process-based
classification of blazars into low- (LSP: LBL+FSRQs), intermediate- (ISP/IBL), and
high-synchrotron-peaked blazars \citep[HSP/HBL;][]{2010ApJ...716...30A,1998MNRAS.299..433F}.

The inferred dominance of the leptonic component from SED modeling of neutrinos
associated blazars e.g. \citet{2018NatAs.tmp..154G} implies
a highly correlated flux variability as well as spectral properties between the
EM bands constituting the two humps in the SED. Thus, for simultaneous continuum
variation indicating emission from the same region, the spectrum of the low-energy
hump is related to that of the high-energy hump. For LBLs/LSPs, to which OJ 287
belongs \citep[e.g.][]{2010ApJ...716...30A}, this implies a direct connection
between the optical-UV (synchrotron) to the MeV-GeV gamma-ray spectrum. The former
being purely of synchrotron origin provides a direct tracer of the underlying particle spectrum. Combining this with the simultaneous
X-ray spectrum, additionally, allows us to explore the extent of the underlying particle
spectrum. This makes the simultaneous optical to X-ray spectrum an excellent observable
to probe the high-energy particle spectrum, free from any artifact/complications that
could affect the high-energy component e.g. multiple IC fields contributing to the
high-energy hump, steepening introduced by the onset of the Klein-Nishina regime
as well as extra-galactic background light etc. The particle spectrum being the
fundamental entity of non-thermal processes has important implications beyond
blazar's high emission, emission region e.g. cosmic rays, astro-particle physics, etc.

Amongst blazars, OJ 287 is one of the best-monitored sources by the transient
observing facility: {\it The Neil Gherels Swift Observatory} because of its frequent 
activity and peculiar properties \citep[and references therein]{2018ApJ...866...11D},
the latter driving most of the coordinated multi-wavelength monitoring \citep[e.g.]
[]{2009PASJ...61.1011S,2017MNRAS.465.4423G,2019AJ....157...95G,2022MNRAS.509.2696S,2017IAUS..324..168K,2021MNRAS.504.5575K}. Studies of simultaneous\footnote{not
strictly, exposure/data-extraction vary significantly in different bands-- order
of a few minutes at optical while sub-hour to hour at X-rays.}  optical-to-X-ray
spectra have reported a strong anti-correlation between optical-UV and X-ray spectral 
index \citep{2017MNRAS.468..426S}. More recently, the source exhibited a very 
different flux and spectral evolution with drastic spectral changes in all the EM 
bands \citep{2017ICRC...35..650B,2018MNRAS.473.1145K, 2018MNRAS.479.1672K,2021ApJ...921...18K,2022MNRAS.509.2696S,2021MNRAS.504.5575K}. The reported spectral phases are unique and OJ 287
is the very first blazar to exhibit spectra that are representative of the entire
blazar class \citep[e.g.][]{2022JApA...43...79K, 2022MNRAS.509.2696S}. Theoretical 
scenarios based modeling of the simultaneous optical to X-ray SED reports both --
a continuation of the optical-UV spectrum (synchrotron) into X-ray bands, making
the X-ray spectrum relatively flatter/softer \citep{2022MNRAS.509.2696S,2020ApJ...890...47P,2001PASJ...53...79I} as well as a sharp steepening of the optical-UV
spectrum at (far) UV energies \citep{2022MNRAS.509.2696S,2009PASJ...61.1011S}.

In the current work, we focus on a detailed spectral investigation of the simultaneous
optical to X-ray spectrum of the observed lowest X-ray flux state
of OJ 287 by the {\it Swift} observatory.
The work is organized as follows. In the next section (\S\ref{sec:data}), we present
the details of the data and reduction procedure. Section \S\ref{sec:results} presents
the analysis and results reporting an optical-UV synchrotron spectrum extending
to X-ray energies and the hardest X-ray spectrum of the source during the lowest
X-ray flux state. Section \S\ref{sec:discuss}
presents the implications of this on particle acceleration scenarios emission
mechanisms with a summary in \S\ref{sec:sum}.

\section{Data Reduction} \label{sec:data}
The {\it Neil Gherels Swift} Observatory \citep{2004ApJ...611.1005G} is a space-based 
facility with three primary payloads -- X-Ray Telescope (XRT; 0.3 -- 10 keV), Ultraviolet Optical Telescope (UVOT), and the Burst Alert Telescope (BAT; 15--150
keV) capable of simultaneously observing optical to hard X-ray band over a wide
range of source brightness state.  In this work, we used all the pointed XRT and
UVOT data of OJ 287\footnote{source too faint for spectral/flux studies at BAT
energies on ID-wise timescale e.g. \citet{2020A&A...637A..55L}.} till December
2022 (MJD: 53510 -- 59940) to perform the spectral study. The dense {\it Swift} 
follow-up since mid-2015 has been carried out by \citet{2017IAUS..324..168K,2021MNRAS.504.5575K}\footnote{the project is now called ``MOMO'' by the PIs}.

{\bf UVOT:} The UVOT data of the source was reduced following the methodology adopted
in \citet{2021ApJ...921...18K}. We used a $\rm 5''$ circular region for the source
and an annular source-free region of inner and outer radii of $\rm 25''$ and $\rm
45''$ respectively for the background, both centred on the source location. We then
executed the \emph{HEASOFT} (v6.30.1) tasks {\it uvotsum} and {\it uvotsource} to
extract the flux density in each filter. The extracted flux densities were subsequently
corrected for reddening using an E(B-V) = 0.0241 with the extinction law of 
\citet{1989ApJ...345..245C}.
    
{\bf XRT:} For the X-ray spectral study, we directly used the scientific grade spectra
and respective ancillary files provided by the UKSSDC\footnote{\url{https://www.swift.ac.uk/user_objects/}}. The underlying pipeline corrects for pile-up in both photon
counting (PC) and window timing (WT) mode data following the procedure detailed in
\citet{2009MNRAS.397.1177E} and subsequent ongoing updates. For OJ 287, none of
the WT mode data requires pile-up treatment (rate $<< 100$ c/s) while PC mode data
having count rate $\gtrsim$ 0.5 c/s were marked for pile-up correction.

To find the best description of the X-ray spectrum, we followed the approach adopted
in \citet{2018MNRAS.479.1672K,2021ApJ...921...18K} using {\it F-test} statistics
to choose the best-fit description between the absorption (tbabs) modified power-law
(PL; $\rm dN/dE \sim E^{-\Gamma}$) and log-parabola (LP; $\rm dN/dE \sim E^{-\alpha-\beta log(E)}$) models. We used a {\it F-test} value of 0.01 below which LP is
preferred and vice-versa. We binned each source spectral file with a minimum of
one count per bin and performed the model fitting (0.3--10 keV) using the $Cash-statistics$ with background (W Statistic in $\rm XSPEC$). During the model fitting, we kept all the
parameters free initially but froze the nH value to the Galactic column density of $\rm 2.4\times 10^{20}~cm^{-2}$ \citep{2016A&A...594A.116H} in the direction of the source, whenever the fit
value was equal or below it. Finally, for spectral study, we corrected all the
extracted X-ray SEDs for nH absorption.

 \begin{figure}
	\includegraphics[width=\columnwidth]{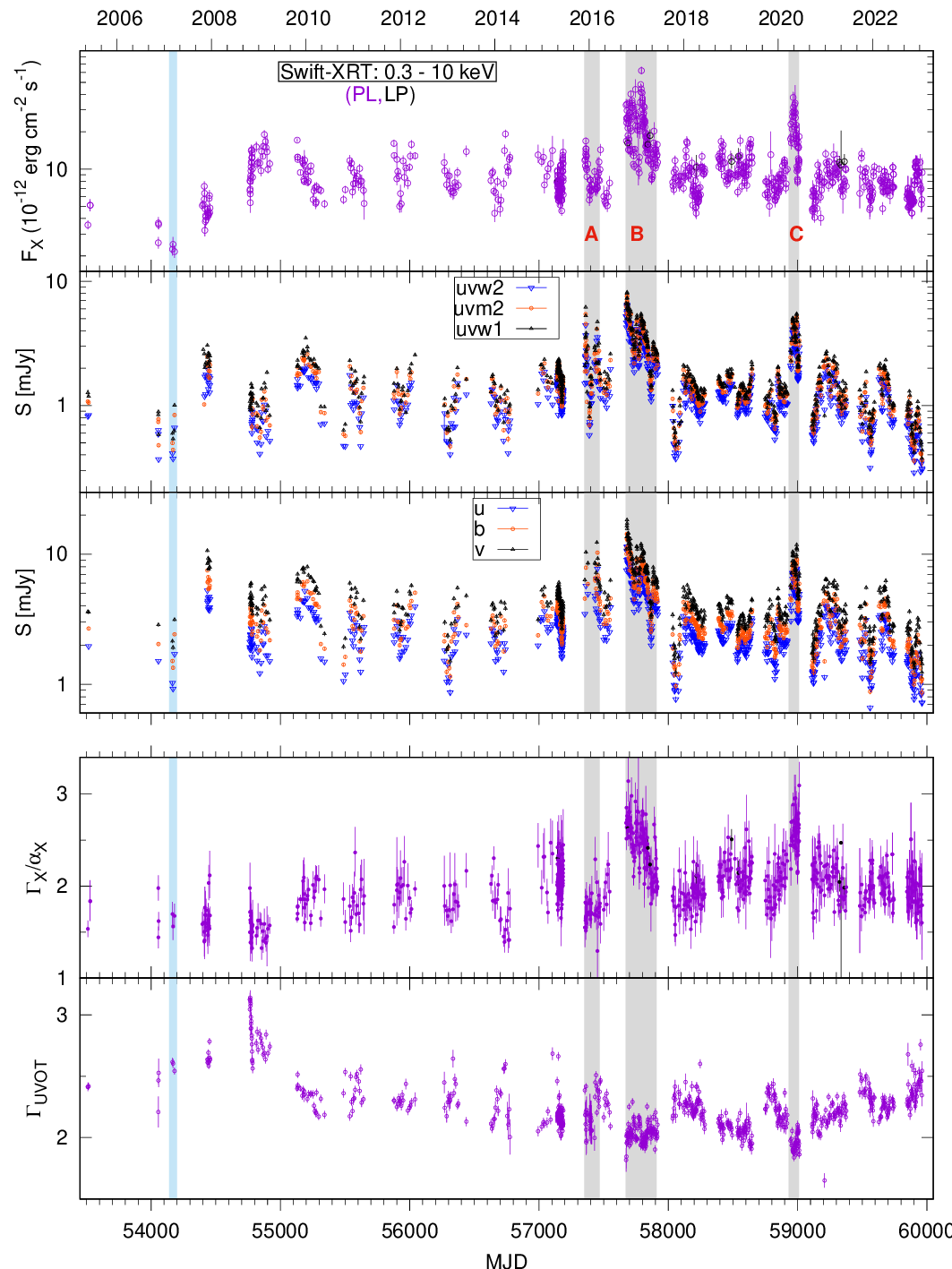}
    \caption{Optical to X-ray light curve of OJ 287 from the {\it Swift}
    facility observation between 2005 to 2022 (ref \S\ref{sec:data}). The spectral
    index from the best fit PL/LP model for X-ray and the PL model of optical-UV 
    is shown in the bottom panels. The cyan band is the focus of this work while the dark-shaded regions
    mark the duration with very different spectral features across the EM bands e.g. A: \citet{2018MNRAS.473.1145K}; B: \citet{2018MNRAS.479.1672K}; C and in-between B-C: \citet{2021ApJ...921...18K}.}
 \label{fig:mwlc}
\end{figure}

\section{SED Analysis and results} \label{sec:results}
The optical to X-ray light curve and spectral behavior extracted from the {\it Swift}
facility are shown in Figure \ref{fig:mwlc}. The UVOT spectral index is extracted
assuming a PL model. The source has undergone a strong flux variation of $> 100$
in X-ray and $\gtrsim 15$ in UVOT bands between the minimum and maximum during this
period. An interesting behavior is the concurrent low-flux state in both optical-UV
and X-ray, marked in the cyan-colored band in  Fig. \ref{fig:mwlc} (MJD $\sim$ 
54160 -- 54180) with the lowest X-ray flux within the cadence, statistics, and
instrument sensitivity. Later, though optical-UV has even gone below this level,
the X-ray has not. Another point to be noted is fewer X-ray points with LP model
in Figure \ref{fig:mwlc} compared to our previous work \citet{2022JApA...43...79K}.
The difference is due to the {\it F-test} probability value used: 0.01 here compared
to 0.05 in \citet{2022JApA...43...79K}. We chose a tighter value because the low-flux 
states --the focus of our study, favor the PL\footnote{relaxing {\it F-test} value will
have no effect on the outcome. It primarily affects the gray shaded part (A, B, C) of
Figure \ref{fig:mwlc}} model.

\begin{figure}
% To include a figure from a file named example.*
% Allowable file formats are eps or ps if compiling using latex
% or pdf, png, jpg if compiling using pdflatex
\includegraphics[width=\columnwidth]{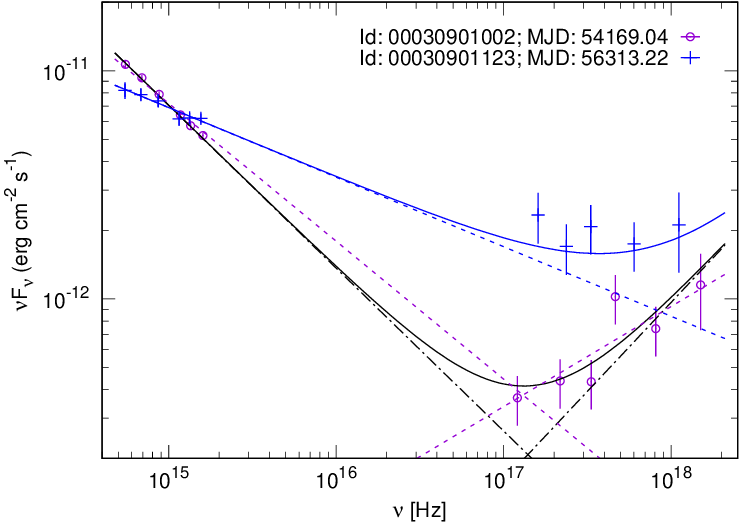}
 \caption{The optical to X-ray SED corresponds to the lowest identified X-ray
 state in purple along with an intermediate state showing a relatively flat X-ray
 spectrum. The purple dashed lines are the best-fit PL model to optical-UV and X-ray
 ($\rm PL:\Gamma_X = 1.5\pm0.1$) separately while the black dot-dashed lines are the
 best-fit from the joint fit of the optical to X-ray
 with the solid black curve representing the overall spectrum. The flat X-ray spectra
 of the intermediate state (blue curve) can be simply reproduced by adding the 
 hard X-ray spectrum from the joint-fit to the corresponding optical-UV synchrotron spectrum.}
 \label{fig:fig2}
\end{figure}

Since our focus is the lowest X-ray flux state and synchrotron is known to affect
the X-ray spectrum \citep{2022MNRAS.509.2696S,2020ApJ...890...47P,2001PASJ...53...79I}, we ordered the IDs first by
X-ray flux (increasing) and then filtered all the IDs with best-fit X-ray
flux within $3\sigma$ of the lowest X-ray flux (resulting in 4 IDs). Among these,
we then looked for the ID with the lowest flux in UVOT bands (ID: 00030901002).
The corresponding optical to X-ray SEDs for this is shown in Figure
\ref{fig:fig2}.

\begin{figure}
 \includegraphics[width=\columnwidth]{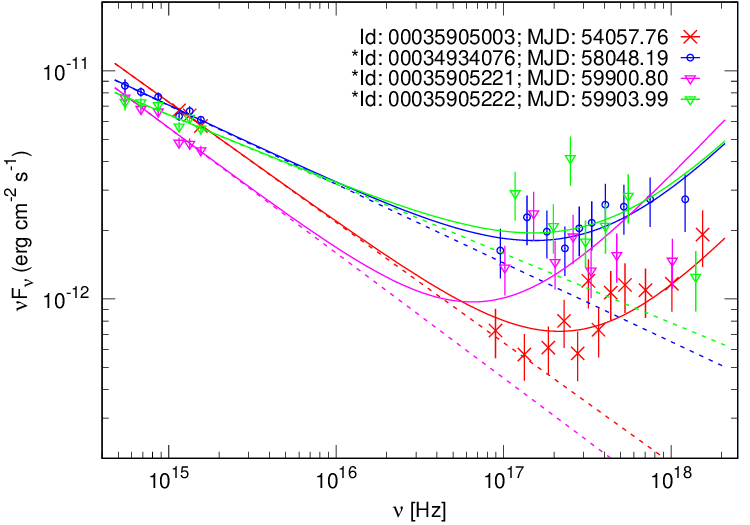}
 \caption{A subset of X-ray spectra, representative of the observed X-ray spectra
 during low and intermediate X-ray flux states, demonstrates that the flat X-ray spectrum
 can be simply reproduced by adding the inferred hard X-ray component to the
 respective optical-UV PL spectrum shown by dashed curves. The ones marked with '*'
 however require variation in normalization (up to a factor of 3 here) of the inferred
 harder X-ray spectrum and in a few cases, a different spectrum.}
 \label{fig:fig3}
\end{figure}

As optical-UV is synchrotron emission, to get the spectral index, we first fitted
a PL model to the UVOT SED using $\chi^2$ statistic in XSPEC. We then extrapolated
the best-fit optical-UV spectrum and examined it vis-a-vis the corresponding X-ray
spectrum. We found that the UVOT PL fit ($\rm \Gamma_{UVOT} = 2.71 \pm 0.03$) extends
well into X-ray. Thus, to get the uncontaminated X-ray spectral parameter, we jointly
modeled the optical to X-ray spectrum with a two-PL model (redden*powerlaw + 
tbabs*powerlaw), for the reddening modified optical-UV PL model contribution to
X-ray energies. We found a harder X-ray photon spectral index of $\rm \Gamma_X = 
1.22 \pm 0.20$ compared to $\rm \Gamma_X \sim 1.5$ when fitting X-ray alone.

To further improve the constraint, we jointly modeled all the observation IDs
(total 7 IDs) having similar X-ray spectral indices (hard; $\Gamma_X: 1.2-1.3$). This
resulted in an improved constraint of $\rm \Gamma_X = 1.29 \pm 0.06$ using 
$\chi^2$-statistic for UVOT and $Cash-statistic$ at X-rays. We re-checked the result
using only $\chi^2$-statistic at both optical-UV and X-ray with X-ray binned to a
minimum of 20 counts per bin. We got a consistent result (within $\rm1\sigma$;
$\rm \Gamma_X = 1.16 \pm 0.08$; $\chi^2/dof\footnote{degree of
freedom}=67.4/61$) but this lacks the lowest X-ray flux state due to insufficient
counts.
% Both the best-fits are consistent within $1\sigma$ ($\chi^2-statistic:\rm \Gamma_X = 1.15 \pm 0.08$)}. 
The best-fit model from the lowest X-ray flux state along with data is shown in Figure \ref{fig:fig2}. This is the
hardest-ever reported X-ray spectrum (0.3 - 10 keV) for OJ 287 to the best of our
knowledge.

We then explored other low and intermediate X-ray flux states, most of which have 
a relatively flatter X-ray spectrum ($\Gamma \sim 1.7 - 2.1$) and found that this
inferred hard X-ray spectrum without or with variation of normalization along with
the respective optical-UV spectrum can naturally explain a majority of the observed
flat X-ray spectrum of the LBL phase of the source. A subset of such IDs, representative
of the different X-ray spectra exhibited by the source are shown in Figures \ref{fig:fig2}
and \ref{fig:fig3}. We also found that a few are indeed different spectrally. These
and the observation-IDs requiring change of normalization of the inferred hard X-ray 
spectrum are marked with '$\rm\ast$' in Figure \ref{fig:fig3}.

\section{Discussion}\label{sec:discuss}
We found an extended optical-UV synchrotron spectrum ($\rm \Gamma_{UVOT} = 2.71
\pm 0.03$), continuing into X-rays during a period with concurrent dip in
optical-UV and X-ray flux (MJD: 54160 -- 54180; cyan band in Fig. \ref{fig:mwlc}), 
corresponding to the lowest reported X-ray flux within the {\it Swift} cadence.
Accounting for this contribution at X-rays, we found a hard X-ray spectrum with photon spectral index $\rm \Gamma_X \sim 1.15 - 1.3$ (0.3-10 keV; ref Fig. \ref{fig:fig2}),
the hardest ever reported from the study of simultaneous optical to X-ray spectrum of
OJ 287. This is in contrast to the bright/active state optical X-ray SEDs of the source
where the X-ray spectrum generally implies a cutoff or sharp steepening of the 
optical-UV synchrotron spectra in the (far)
UV region \citep[e.g.][]{2022MNRAS.509.2696S,2009PASJ...61.1011S}. This X-ray spectrum is
consistent with the spectrum reported in the hard X-ray band from {\it Swift}-BAT
measurement \citep[50-300 keV;][]{2020A&A...637A..55L}. We further found that a 
majority of the flat
X-ray spectra during the low and intermediate X-ray flux states can be naturally
explained by simply adding the inferred hard X-ray spectrum with the respective
optical-UV  spectrum either originally or with a variation of the normalization (ref Figs. \ref{fig:fig2} and \ref{fig:fig3}).  A few cases, however, indeed have different
spectrum (refer to Fig. \ref{fig:fig3}). 
% A detailed investigation will be presented in the follow-up work \citep[in prep.]{2023KushPrep}.

Blazar optical continuum is well-understood to be synchrotron and thus, the observed
spectrum directly traces the underlying particle spectrum. The same holds for the
IC spectrum i.e. X-ray, if its away from the peak and IC happens in the Thomson
regime. For a power-law particle distribution of spectral index $p$ ($\rm N(E) \sim
E^{-p}$), the observed radiation energy spectrum is $\rm f \sim E_{ph}^{-(p-1)/2}$.
OJ 287 being an LBL source, the X-ray is a good tracer of the low-energy part while
the optical-UV directly traces the high-energy part of the broken power-law particle distribution required for broadband SED modeling \citep[e.g.][]{2022MNRAS.509.2696S}.
The reported optical to X-ray spectrum corresponding to the lowest X-ray flux thus
implies a hard low-energy particle spectrum of $\rm 1.3 - 1.6$ and a high-energy
spectrum of $\rm \sim 4.4$.

For blazar's synchrotron spectrum, the observed frequency is related to the rest
frame via $\rm \nu_{obs} = \delta/(1+z) \gamma^2 \nu_L $ where $\rm \nu_L$ is the
Larmor frequency, $\rm \gamma$ is the electron Lorentz factor, and $\rm \delta$ is
the bulk Doppler factor. Further, $\rm \nu_L = eB/2\pi m_e c$, depends solely on
the magnetic field, B. Thus for a given B and $\rm \delta$, the extent of the spectrum 
depends on the Lorentz factor (energies) of the particles. Since X-ray lies at the
tail of the high-energy synchrotron spectrum, the combined optical to X-ray spectrum
thus provides a potential direct tracer of the extent of the high-energy particle
spectrum. The finding of an extended optical-UV synchrotron spectrum reported here
clearly indicates a much extended high-energy particle spectrum compared to the
bright X-ray phases of the source. On the other hand, the natural explanation of
most of the flat X-ray spectra associated with the low and intermediate X-ray flux
states of the LBL/LSP state of the source by adding the
high-energy synchrotron tail to the lowest hard X-ray spectrum further strengthens
previous claims/finding that most of the X-ray spectral changes are due to optical-UV
synchrotron spectrum \citep{2022MNRAS.509.2696S}.

The relativistic particle spectrum has two main ingredients -- the spectral index
and the extent (energy-limits) of the spectrum. The spectral index is related
to the particle acceleration processes and the extent of the spectrum is related
to the size of the acceleration region -- the Hilas energy criteria \citep{1984ARA&A..22..425H}. The reported extended high-energy spectrum implies a comparatively
larger acceleration/emission region compared to the flaring states that in general
indicate a steepening or cutoff. The inference is also consistent with short variability timescales, indicating a compact emission region during flares. Indirectly the
larger acceleration/emission region could possibly
be the large-scale jet emission. Since the reported hard X-ray is sum of synchrotron tail and IC -- mostly of the low-energy part of particle distribution, a harder
spectrum at X-ray may not be a direct reflection of harder particle spectrum but 
can also result from a higher-lower-energy cutoff of the underlying particle
spectrum \citep{1996ApJ...463..555I} -- consistent with a larger acceleration/emisison
region argued above. But if its actual then the lower-limit of
the hard particle spectrum indicates magnetic reconnection as the most likely
driver \citep[e.g.][]{2014PhRvL.113o5005G} while the upper-limit appears consistent
with the shock scenario \citep{2001RPPh...64..429M}.

Though OJ 287 has been studied extensively at all wavelengths, and especially
at optical and X-rays, most of the studies focusing on the latter have separately
explored optical-UV and X-rays \citep[e.g.][and references therein]{2017MNRAS.468..426S,2021ApJ...921...18K,2022MNRAS.513.3165K} or the X-ray flux have been comparatively
higher \citep[e.g.][]{2022MNRAS.509.2696S,2009PASJ...61.1011S}. Given the blazar
broadband emission, a consideration of all likely processes is a must to gain further insight into the source behavior and thus, the jet physics, extra/hidden emission components, etc. A detailed outcomes of this systematic analysis will be will be
presented in the accompanying paper \citep[in preparation]{2023KushPrep}.
    
\section{Summary and Conclusions}\label{sec:sum}
We explored the simultaneous optical to X-ray observations (spectra) of the BL
Lacertae object OJ 287 by the {\it Neils Geherel's Swift Observatory} to date with
a focus on the lowest X-ray state and found a power-law optical-UV synchrotron
continuum of photon spectral index $\rm \Gamma_{UVOT} = 2.71 \pm 0.03$ that
extends to X-ray energies. Modeling the corresponding X-ray spectrum accounting
synchrotron component contribution, we found a hard 0.3-10 keV X-ray spectrum
with $\rm \Gamma_X \sim 1.15 - 1.3$. This is harder than the reported hardest
spectrum ($\rm \Gamma_X \sim 1.5$) considering only the X-ray data and is consistent
with the hard X-ray spectrum reported from the {\it Swift}-BAT data (20-100 keV).
The optical-UV directly traces the underlying high-energy particle spectrum
with a PL spectrum of $\sim 4.4$ while X-ray implies a harder PL spectrum of
$1.3-1.6$. The harder spectrum in the latter case may not be real and could result
from a comparatively higher lower-energy-cutoff in the particle spectrum.

The reported hard X-ray can naturally explain most of the observed flat X-ray spectra 
when combined with the corresponding optical-UV spectrum, 
% though reproduction of some spectra demands variation of the normalization of this component. 
The finding further strengthens previous claim that most of the flat X-ray spectra
of the source in the LBL/LSP spectral phase of the source are due to the optical-UV synchrotron spectrum extending to X-ray energies.

The extended optical-UV spectrum implies an extended high-energy particle spectrum,
and thus, a relatively large acceleration/emission region size per the Hilas criteria
and is consistent also with the indication of a comparatively higher lower-energy-cutoff
in the particle spectrum implied from the hard X-ray spectrum.  The larger size
suggests the inferred low-emission component could possibly be associated with the
large-scale jet.

\section*{Acknowledgments}
The author thank the anonymous referees for valuable inputs and suggestions.
The author acknowledges financial support from the Department of Science and Technology (DST), Government of India, through the DST-INSPIRE faculty grant (DST/
INSPIRE/04/2020/002586). This work made use of data supplied by the UK Swift
Science Data Centre at the University of Leicester.
\\
{\it Facilities:} Swift. \\
\\
{\it Software:} {Gnuplot (version: 5.2; \url{http://www.gnuplot.info/}), SHERPA,
HEASOFT}

%\software{}

% \begin{thebibliography}{oj287_uvX_v11.bib}
%  
% \end{thebibliography}

\bibliographystyle{aasjournal}
\bibliography{oj287_uvX_v12.bib}

\end{document}